# An Efficient Method for Multi-Parameter Mapping in Quantitative MRI using B-Spline Interpolation

Willem van Valenberg, Stefan Klein, Frans M. Vos, Kirsten Koolstra, Lucas J. van Vliet, Dirk H.J. Poot

*Abstract* — **Quantitative MRI methods that estimate multiple physical parameters simultaneously often require the fitting of a computational complex signal model defined through the Bloch equations. Repeated Bloch simulations can be avoided by matching the measured signal with a precomputed signal dictionary on a discrete parameter grid (i.e. lookup table) as used in MR Fingerprinting. However, accurate estimation requires discretizing each parameter with a high resolution and consequently high computational and memory costs for dictionary generation, storage, and matching.**

**Here, we reduce the required parameter resolution by approximating the signal between grid points through B-spline interpolation. The interpolant and its gradient are evaluated efficiently which enables a least-squares fitting method for parameter mapping. The resolution of each parameter was minimized while obtaining a user-specified interpolation accuracy. The method was evaluated by phantom and in-vivo experiments using fully-sampled and undersampled unbalanced (FISP) MR fingerprinting acquisitions. Bloch simulations incorporated relaxation effects $(T_1, T_2)$, proton density $(PD)$, receiver phase $(\varphi_0)$, transmit field inhomogeneity $(B_1^+)$, and slice profile. Parameter maps were compared with those obtained from dictionary matching, where the parameter resolution was chosen to obtain similar signal (interpolation) accuracy. For both the phantom and the in-vivo acquisition, the proposed method approximated the parameter maps obtained through dictionary matching while reducing the parameter resolution in each dimension $(T_1, T_2, B_1^+)$ by – on average – an order of magnitude. In effect, the applied dictionary was reduced from $1.47$ GB to $464$ KB. Furthermore, the proposed method was equally robust against undersampling artifacts as dictionary matching. Dictionary fitting with B-spline interpolation reduces the computational and memory costs of dictionary-based methods and is therefore a promising method for multi-parametric mapping.**

*Index Terms*—B-spline interpolation, dimensionality reduction, least-squares minimization, magnetic resonance fingerprinting, singular value decomposition, quantitative magnetic resonance imaging

## I. Introduction

QUANTITATIVE MRI (qMRI) methods measure the magnetic properties of tissues, described by parameters such as relaxation times $(T_1, T_2, T_2^*)$ and proton density $(PD)$. Many of these methods require knowledge of inhomogeneities in the static $(\Delta B_0)$ and/or transmit $(B_1^+)$ magnetic field in order to obtain accurate parameter maps. Changes in the magnetic properties of tissues have been linked to various pathologies [1].

Magnetic resonance fingerprinting (MRF) is a recently introduced paradigm to acquire multiple parameters within a short scan time [2]. MRF methods use a pulse sequence with varying flip angles and repetition times to acquire images with many different contrasts. In each voxel, the signal's time course is assumed to be specific to the parameter combination representing the underlying tissue. Usually, each contrast is undersampled, but by varying the k-space trajectory among the contrasts, undersampling artifacts are assumed separable from the true signal of a voxel. Parameter estimation is done by matching the acquired signal course in a voxel to a dictionary that contains the simulated signals, or atoms, for a grid of parameter combinations (e.g. $T_1, T_2, T_2^*, \Delta B_0, B_1^+$). This matching avoids the fitting of an explicit signal model, which in MRF would require repeatedly solving Bloch equations, which is computationally expensive. Before MRF, dictionary

†This work was financially supported by the Netherlands Organization for Health Research and Development ZonMW (project number 104 003 005), and the Dutch Technology Foundation STW (CARISMA 11631).

W. van Valenberg is with the Department of Imaging Physics, Delft University of Technology, 2628 CJ Delft, The Netherlands, and with the Departments of Medical Informatics and Radiology, Erasmus MC, 3015 GE Rotterdam, The Netherlands (e-mail: w.vanvalenberg@tudelft.nl).

S. Klein is with the Departments of Medical Informatics and Radiology, Erasmus MC, 3015 GE Rotterdam, The Netherlands (e-mail: s.klein@erasmusmc.nl).

F. M. Vos is with the Department of Imaging Physics, Delft University of Technology, 2628 CJ Delft, The Netherlands, and with the Department of Radiology, Amsterdam UMC, 1109 AZ Amsterdam, The Netherlands (e-mail: f.m.vos@tudelft.nl).

K. Koolstra is with the Department of Radiology, Leiden University Medical Center, 2333 ZA Leiden, The Netherlands (e-mail: k.koolstra@lumc.nl).

L. J. van Vliet is with the Department of Imaging Physics, Delft University of Technology, 2628 CJ Delft, The Netherlands (e-mail: l.j.vanvliet@tudelft.nl).

D. H. J. Poot is with the Department of Imaging Physics, Delft University of Technology, 2628 CJ Delft, The Netherlands, and with the Departments of Medical Informatics and Radiology, Erasmus MC, 3015 GE Rotterdam, The Netherlands (e-mail: d.poot@erasmusmc.nl).

matching was also applied in other qMRI methods in order to improve $T_1$ estimation [3], [4], $T_2$ estimation [5], and water/fat separation [6], [7].

The original MRF paper obtained $T_1, T_2, \Delta B_0$, and $PD$ parameter maps with a pseudorandom 2D inversion-recovery balanced steady state free-precession (IR-bSSFP) sequence with variable-density-spiral readout [2]. Subsequently, a modification to the original scheme was proposed by applying the fast-imaging with steady-state precession (FISP) sequence, which includes unbalanced gradients [8]. This reduced the influence of static field inhomogeneities at the cost of a lower signal-to-noise ratio (SNR). In recent years, these MRF methods have been extended to measure other properties of interest such as diffusion [9], perfusion [10], and chemical exchange [11]. Furthermore, including transmit field inhomogeneity ($B_1^+$) and slice profile in the fitting has been shown to increase the accuracy of the resulting relaxometry maps [12], [13].

However, dictionary matching becomes problematic when the number of estimated parameters is increased. The number of atoms increases exponentially with the number of parameters, and consequently also the computational and memory costs of generating, storing, and matching to the dictionary. This is especially prohibitive if the required precision for each parameter is high since this requires many steps along each dimension of the dictionary.

Several clever strategies were introduced to reduce the computational and memory demands of large dictionaries. Smarter search strategies can significantly reduce the matching time [14], but the size of the dictionary is limited by the available memory. The atoms can be compressed with a singular value decomposition (SVD) to lower the computational and memory costs for matching and storing the dictionary [15]. However, using too few singular vectors degrades the results. More recent work proposed interpolating the signal with a polynomial hyperplane fitted on a sparsely sampled dictionary [16]. However, this method was applied to a two-parameter case only ($T_1$ and $T_2$) and the accuracy of the parameter maps remained limited to an a priori defined refinement factor.

We propose parameter estimation by fitting the acquired MR signal with a continuous signal model defined through B-spline interpolation of a sparse dictionary. The interpolation targets to maintain the estimation accuracy while reducing the resolution of each parameter and consequently the computational and memory costs of the dictionary. This would enable the estimation of an increased number of parameters simultaneously. B-spline interpolation is commonly used in image interpolation because it is a flexible and efficient technique that has minimal support for a desired interpolation error [17], [18]. In particular, the derivative of the interpolant can be calculated efficiently [18] which allows a gradient-based optimization technique for fitting a measured signal to the dictionary. Additionally, we introduce a method to estimate the parameter resolution in the dictionary that is required to achieve a user-specified interpolation accuracy. The proposed method is evaluated on simulated data and measurements taken from phantom and in-vivo experiments. The efficiency of our dictionary fitting framework is compared to matching with a dictionary of equal accuracy. We hypothesize that the proposed method accurately estimates a comprehensive set of parameters based on a significantly smaller dictionary.

II. METHODS

A. *Parameter Estimation*

A general qMRI method measures the complex-valued signal $\boldsymbol{m}$ of a voxel at $M$ time points. The signal is assumed to be a function of $P$ parameters $\boldsymbol{\theta} = [\theta_1, \ldots, \theta_P] \in \Theta \subseteq \mathbb{R}^P$, contaminated by Gaussian noise:

$$\boldsymbol{m} = \rho \boldsymbol{s}(\boldsymbol{\theta}) + \boldsymbol{\sigma} \qquad (1)$$

The signal model $\boldsymbol{s}(\boldsymbol{\theta}) \in \mathbb{C}^M$ is the pulse sequence specific solution of the Bloch equations, and the scaling factor $\rho \in \mathbb{C}$ is dependent on the proton density and the receiver sensitivity. Note that we assume a single-compartment model in each voxel, so that $\rho$ is a single complex number. The Gaussian noise $\boldsymbol{\sigma} \in \mathbb{C}^M$ is considered identical and uncorrelated between measurements and receiver channels.

Parameter estimation is often done by least-squares fitting:

$$[\hat{\boldsymbol{\theta}}, \hat{\rho}] = \arg\min_{\boldsymbol{\theta} \in \Theta, \rho \in \mathbb{C}} \|\boldsymbol{m} - \rho \boldsymbol{s}(\boldsymbol{\theta})\|_2^2 \qquad (2)$$

However, the signal model $\boldsymbol{s}(\boldsymbol{\theta})$ is computationally complex for MRF, since the signal at a given time point depends on the signal's history during previous steps. As such, it requires solving the Bloch equations step-by-step. This makes conventional optimization techniques for solving Eq. 2 expensive.

B. *Dictionary Matching*

MRF avoids repeated evaluation of $\boldsymbol{s}(\boldsymbol{\theta})$ by matching the acquired signal to a precomputed dictionary, i.e. signals on a discrete grid of parameter values [2]. The dictionary atom with index $\boldsymbol{k} \in \mathbb{N}^P$ corresponds to parameter values $\boldsymbol{\theta} = \boldsymbol{f}(\boldsymbol{k})$, where the mapping $\boldsymbol{f}(\boldsymbol{v})$ is defined for continuous grid position $\boldsymbol{v} \in \mathbb{R}^P$ in order to facilitate interpolation (see below).

The dictionary matching step in MRF finds for a measured signal $\boldsymbol{m}$ the grid point $\hat{\boldsymbol{k}}$ and consequently the associated parameter combination $\hat{\boldsymbol{\theta}} = \boldsymbol{f}(\hat{\boldsymbol{k}})$ by

$$\hat{\boldsymbol{k}} = \arg\max_{\boldsymbol{k}} |\boldsymbol{m}^H \boldsymbol{s}(\boldsymbol{f}(\boldsymbol{k}))| / \|\boldsymbol{s}(\boldsymbol{f}(\boldsymbol{k}))\|_2. \qquad (3)$$

The superscript $H$ indicates the Hermitian conjugate. The complex scaling factor $\hat{\rho}$ is subsequently determined through the least-square solution:

$$\hat{\rho} = \left(\boldsymbol{s}(\hat{\boldsymbol{\theta}})^H \boldsymbol{m}\right) / \left(\boldsymbol{s}(\hat{\boldsymbol{\theta}})^H \boldsymbol{s}(\hat{\boldsymbol{\theta}})\right) \in \mathbb{C} \qquad (4)$$

The solution $[\boldsymbol{f}(\hat{\boldsymbol{k}}), \hat{\rho}]$ of Eqs. 3 and 4 is also the solution of

Eq. 2 when cast as a discrete optimization problem over the parameter values $\boldsymbol{\theta} = f(\boldsymbol{k})$ (see Supplementary Materials A).

The number of dictionary atoms increases linearly with the number of discretized values ($K_p$) of each parameter and exponentially with the number of parameters ($P$). Therefore, high-precision multi-parameter maps are computationally infeasible since the computational and memory cost for dictionary generation, storage, and matching scale linearly with the number of dictionary atoms. Singular value decomposition (SVD) can alleviate these effects by projecting both the measurement data $\boldsymbol{m}$ and the dictionary atoms $\boldsymbol{s}(f(\boldsymbol{k}))$ to a lower dimensional space:

$$\boldsymbol{m}_L = \boldsymbol{V}_L^T \boldsymbol{m} \in \mathbb{C}^L$$
$$\boldsymbol{s}_L(f(\boldsymbol{k})) = \boldsymbol{V}_L^T \boldsymbol{s}(f(\boldsymbol{k})) \in \mathbb{C}^L \quad (5)$$

where $\boldsymbol{V}_L^T$ contains the singular vectors corresponding to the $L$ largest singular values. As a result, the memory costs of storing the dictionary and the cost of the dictionary-matching step in Eq. 3 reduces by a factor $L/M$ [14]. However, the results degrade when using too few singular vectors and multi-parameter mapping is still computationally demanding since the SVD only reduces the number of time points and not the number of atoms.

*C. Dictionary Fitting*

To enhance the precision of the parameter maps while limiting the number of grid points, we propose a dictionary fitting[1] framework in which the signal is modelled on the whole, continuous parameter domain through interpolation of a sparsely sampled dictionary. We define the B-spline interpolated signal of order $n$ at grid position $\boldsymbol{v} \in \mathbb{R}^P$ (without SVD) as [17]:

$$\tilde{\boldsymbol{s}}(\boldsymbol{v}) = \sum_{\boldsymbol{k} \in \mathbb{N}^P} \boldsymbol{c}(\boldsymbol{k}) \beta^n(\boldsymbol{v} - \boldsymbol{k}) \quad (6)$$

Here, $\boldsymbol{c}(\boldsymbol{k}) \in \mathbb{C}^M$ indicates the B-spline coefficient for each dictionary atom and $\beta^n(\boldsymbol{v})$ is the product of B-spline basis functions of order $n$ along each dimension:

$$\beta^n(\boldsymbol{v}) = \prod_{p=1}^{P} \beta^n(v_p) \quad (7)$$

For details on B-spline interpolation, including the exact definition of the B-spline basis functions $\beta^n(\boldsymbol{v})$ we refer to a general background paper [17]. Essentially, the $n$th order B-spline basis function is a piecewise polynomial of degree $n$ with width of support $n + 1$. The coefficients $\boldsymbol{c}(\boldsymbol{k})$ can be obtained via a closed-form solution, such that $\tilde{\boldsymbol{s}}(\boldsymbol{k}) = \boldsymbol{s}(f(\boldsymbol{k}))$. In effect, the interpolated function intersects the dictionary atoms exactly, while there is continuity up to the nth derivative.

With SVD compression, the B-spline interpolated signal $\tilde{\boldsymbol{s}}_L(\boldsymbol{k}) \in \mathbb{C}^L$ and its coefficients $\boldsymbol{c}_L(\boldsymbol{k}) \in \mathbb{C}^L$ are defined by replacing $\boldsymbol{s}(f(\boldsymbol{k})) \in \mathbb{C}^M$ with $\boldsymbol{s}_L(f(\boldsymbol{k})) \in \mathbb{C}^L$. Through recursive implementation of the spline interpolation [18], the computational cost of evaluating both $\tilde{\boldsymbol{s}}_L(\boldsymbol{v})$ and its gradient is only $\mathcal{O}(Ln^P)$. The incorporation of B-spline interpolation and SVD compression in Eq. 2 yields:

$$[\hat{\boldsymbol{v}}, \hat{\rho}] = \arg \min_{\boldsymbol{v}, \tilde{\rho}} \|\boldsymbol{m}_L - \rho \tilde{\boldsymbol{s}}_L(\boldsymbol{v})\|_2 \quad (8)$$

The optimization problem in Eq. 8 is solved using the `fmincon` routine from MATLAB (The Mathworks, Natick, MA) with the `trust-region-reflective` algorithm. Dictionary matching determines the initial value and the optimization stops when the error reduction is below $10^{-5}$ in subsequent steps or after 100 iterations. Subsequently, we set $\hat{\boldsymbol{\theta}} = f(\hat{\boldsymbol{v}})$, and the proton density ($PD$) and receiver phase ($\varphi_0$) are determined by the modulus and phase of the complex scaling factor $\hat{\rho}$. The accuracy of the parameter estimates from Eq. 8 depends on the invertibility of the forward model $\boldsymbol{s}(\boldsymbol{\theta})$ (i.e the applied acquisition), and on the approximation errors due to the SVD projection (Eq. 5) and the B-spline interpolation (Eq. 6) of which the latter is investigated in the following section.

*D. Parameter Resolution*

The interpolation error over the range $\Theta_p$ of parameter $\theta_p$ decays as $\mathcal{O}\left((1/K_p)^{n+1}\right)$ for B-spline order $n$ and number of discretized values $K_p$ [17]. The interpolation error at a specific position $\boldsymbol{v} \in \mathbb{R}^P$ is defined by

$$E_{int}(\boldsymbol{v}) := \|\tilde{\boldsymbol{s}}(\boldsymbol{v}) - \boldsymbol{s}(f(\boldsymbol{v}))\|_2. \quad (9)$$

To reduce the computation and memory costs of the dictionary, we aim to find for each spline order the smallest number of atoms such that $E_{int}(\boldsymbol{v})$ is below a user-specified threshold $\alpha$ for all $\boldsymbol{v}$ with $f(\boldsymbol{v}) \in \Theta$. We set the parameter resolution of the dictionary based on the interpolation error on the boundary of $\Theta$, where we assume the error is maximal. So the parameter resolution of the dictionary is determined under the assumption that the interpolation error is maximal at the boundary of $\Theta$. Consequently, the number of atoms ($K_p$) in parameter domain $\Theta_p$ is estimated based on the interpolation error along $2^{P-1}$ edges where the other parameters obtain their maximal/minimal value. On each edge, we define $\tilde{\boldsymbol{s}}(\boldsymbol{v})$ through interpolation of increasing number of atoms $K_p = 2^{j-1} + 1$ uniformly sampled on the grid, starting with $j = 1$ (i.e. the minimum and maximum of parameter $\theta_p$) until a user-specified maximum $J$. For each iteration $j$ and spline order $n$, we determine the overall interpolation error as the maximum of $E_{int}(\boldsymbol{v})$ evaluated at the midpoints between atoms on each edge. We select the $K_p$ for which the overall interpolation error is below a chosen value $\alpha$ for the given number and all further refinements. We include values $K_p \neq 2^j + 1$ in this selection by estimating the overall interpolation error between the $J$ iterations through linear interpolation.

---
[1] We use the term *dictionary matching* for the discrete optimization in Eq. 3 and *dictionary fitting* for the continuous optimization in Eq. 8.

*E. Dictionary Design*

The dictionary-fitting framework is tested with a FISP MRF pulse sequence [8]. The generated dictionary contains the simulated signals as a function of $P = 3$ parameters: longitudinal relaxation time $T_1 \in [5, 6000]$ ms, transversal relaxation time $T_2 \in [5, 2000]$ ms, and transmit field inhomogeneity $B_1^+ \in [0.5, 1.5]$. Thus $\boldsymbol{\theta} = (T_1, T_2, B_1^+) \in \mathbb{R}^3$. We define $\boldsymbol{\theta} = \boldsymbol{f}(\boldsymbol{v}) = [f_1(v_1), f_2(v_2), f_3(v_3)]$, where $f_p$ maps $[1, K_p]$ to $\Omega_p$ logarithmically for $T_1$ and $T_2$, and linearly for $B_1^+$. This choice was made since the signal amplitude has a known exponential dependence on $T_1$ and $T_2$. For B-spline orders $n \geq 2$, we avoid interpolation issues near the boundary by extending the grid with one position outside of $\Theta$ and set the derivative of the interpolant equal the numerical derivative as boundary condition.

The pulse sequence was modelled with an event-based approach with RF pulses, gradient pulses, and signal readout at specified time points. Adiabatic inversion pulse and gradient pulses were modelled as instantaneous rotations. The slice profile was modelled through 10,000 spins that were uniformly distributed over twice the slice width (FWHM). To reduce computational complexity, the true excitation pulse was replaced by a pulse consisting of 7 time steps with amplitude, phase, and duration of each step optimized to approximate the true response of a 90 degree pulse without relaxation. This approximation had a relative error below 1% (with $L^2$ norm) and reduced the computational complexity by a factor of 14 compare to applying the full RF pulse. The simulated signal $\boldsymbol{s}(\boldsymbol{\theta})$ has the maximal amplitude of 1 when all spins are coherent in the transverse plane.

III. EXPERIMENTS

The proposed approach was evaluated on simulated, phantom, and in-vivo data. In each experiment, we used 1000 flip angles and repetition times as specified in the original FISP MRF article [8]. Other settings were: inversion time $T_I = 40$ ms, echo time $T_E = 2.5$ ms, and delay $T_D = 5000$ ms after each pulse train. Excitation pulses had a duration of 1 ms, a time-bandwidth product of 3, and a slice width of 5 mm.

The code of the dictionary fitting framework and the experiments performed is provided for reference purposes at https://bitbucket.org/bigr_erasmusmc/dictionary_fitting. All processing was done in MATLAB using a single 2.1 GHz core (AMD Opteron 6172).

*A. Dictionary Design and Generation*

The resolution of the parameters in the dictionary was estimated for each combination of pulse sequence and spline order as described in Section II.D. We set the interpolation error threshold to $\alpha = 5 \cdot 10^{-4}$, which is below the noise level observed in our practical experiments, and the maximum number of iterations $J = 10$, since higher number of atoms were computationally infeasible. The total dictionary size was calculated as the product of required number of atoms for each parameter to pass the target error.

Two dictionaries were generated in order to evaluate the proposed method. Dictionary fitting (Eq. 8) used a sparse dictionary based on the parameter resolution prescribed for second ($n = 2$) order B-spline interpolation. As a reference, parameter estimation was done through dictionary matching (Eq. 3) using a dense dictionary with parameter resolution prescribed for zeroth ($n = 0$) order B-spline interpolation (i.e. nearest neighbor).

*B. Dictionary Evaluation*

We evaluate if the interpolation error in the interior is below the prescribed threshold $\alpha$ in the dense and sparse dictionary with respectively zeroth and second order B-spline interpolation. The interpolation error was determined by Eq. 9 at a 1000 positions $\boldsymbol{v}$, sampled uniformly between 1 and $K_p$ for each dimension $p$, with condition that $f_2(v_2) = T_2 \leq T_1 = f_1(v_1)$ to ensure physically realistic values. This validation of the interpolation accuracy was performed without SVD compression in order to separate different sources of error.

*C. Phantom and In-vivo Experiment*

Practically, we evaluated the dictionary-fitting framework on a 3T Ingenia scanner with a 32-channel head coil (Philips Healthcare, Best, The Netherlands) on a phantom and a healthy volunteer. Data sampling was done using a spiral trajectory that was rotated 7.5 degrees between samples and required 48 interleaves to fully sample a $128 \times 128$ matrix.

Parameter maps were determined by dictionary fitting (Eq. 8) and matching (Eq. 3) with respectively the sparse and dense dictionaries (see Sec. II.D) with SVD compression. The number of singular values $L$ was set to 30 which is in accordance with previous work [16]. The effect of the compression was evaluated on the parameter maps obtained from the fully-sampled in-vivo experiment.

For the phantom experiment we used the NIST system phantom that contains contrast spheres with calibrated $T_1$ and $T_2$ values [19]. We reconstructed images based on an undersampled (1 interleave) and fully sampled (48 interleaves) acquisition. The total scan time was 18 and 871 seconds for the undersampled and fully sampled acquisition, respectively. The accuracy of the methods was compared by a paired t-test of the mean estimated $T_1$ and $T_2$ values in each contrast sphere for both the fully sampled and undersampled data. To quantify the efficiency of our method, we recorded the computation time and memory usage for the dictionary calculation, storage and fitting.

The in-vivo experiment concerned acquiring a 2D slice of the brain of a healthy volunteer. The study was approved by the LUMC review board for Medical Ethics and the volunteer gave an informed consent. Initially, we compare the parameter maps obtained with dictionary matching and fitting from reconstructed images of the fully sampled (48 interleaves) acquisition. Subsequently, we retrospectively undersampled the k-spaces of the fully sampled acquisition by selecting 1, 2, 4, 6, 12, 24, and 48 interleaves. Image reconstruction based on the selected interleaves was performed by a non-uniform Fourier transform with density compensation. For each number of interleaves, we determined the $T_1$ and $T_2$ maps obtained through dictionary matching and fitting, and compared those with the maps from the fully sampled data.

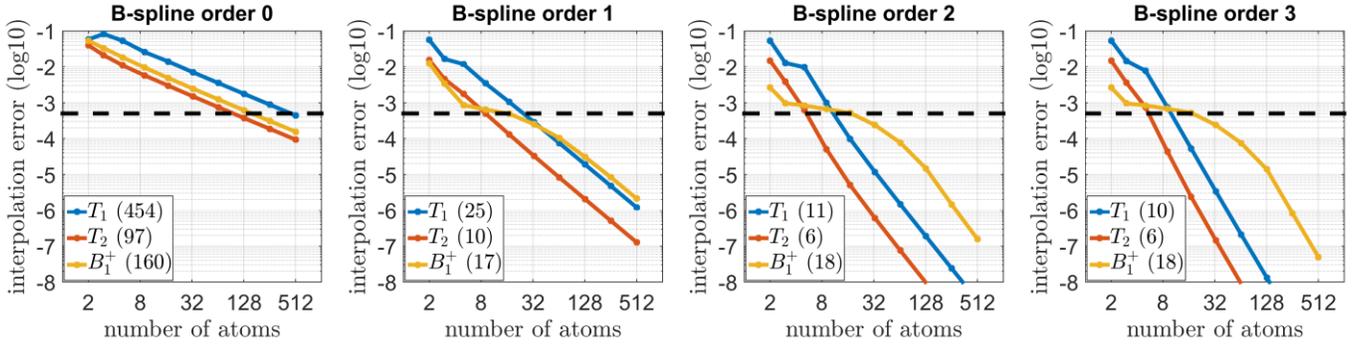

Fig. 1. Interpolation error on the edge of parameter space as function of number of atoms in each dimension. Dashed line indicates target error and the legend shows for each parameter the minimum number of atoms required to obtain the target error. The number of atoms in each parameter dimension reduces approximately an order of magnitude between zeroth and higher order B-spline interpolation.

## IV. RESULTS

### A. Dictionary Design and Generation

Fig. 1 shows the predicted interpolation error as a function of the number of atoms in each parameter dimension ($K_p$) for spline orders $n = \{0,1,2,3\}$. The interpolation error was quantified by the maximum value of $E_{int}(v)$ over the midpoints between sampled positions. The legend reports the minimum number of atoms required for $E_{int}(v) < 5 \cdot 10^{-4}$ (which excludes the boundary padding for $n \geq 2$). The method predicts for zeroth order B-spline interpolation that the target interpolation error is achieved using $7.05 \cdot 10^6$ atoms (i.e. $454 \cdot 97 \cdot 160$). With second order spline, the total number of atoms including the boundary dropped to 2080 (i.e. $13 \cdot 8 \cdot 20$), a factor of $3.38 \cdot 10^3$ reduction.

Dictionaries with both these parameter resolutions were generated. To do so the average computation time of a single atom based on the Bloch simulation was 6.20 seconds. SVD compression to 30 vectors reduced the memory cost of the dense dictionary from 48.0 GB to 1.47 GB, and of the sparse dictionary from 14.6 MB to 464 KB.

### B. Dictionary Evaluation

Fig. 2 shows the interpolation error $E_{int}(v)$ of a 1000 positions in the interior for the dense and sparse dictionaries with respectively zeroth and second order B-spline interpolation (without SVD). The interpolation errors are shown as function of $T_1$, $T_2$, and $B_1^+$. Note that the constraint $T_1 \geq T_2$ biased sampled positions to high $T_1$ and low $T_2$ values. The root-mean-square value of all interpolation errors was $4.1 \cdot 10^{-4}$ and $2.8 \cdot 10^{-4}$ for respectively the dense and sparse dictionary, with maxima of $31 \cdot 10^{-4}$ and $16 \cdot 10^{-4}$. The interpolation error was above the target error for 15.5% of the sampled positions with dictionary matching, and for 7.0% of the same sampled positions with dictionary fitting. It can be observed that in particular the interpolation error with second-order B-splines was highest for test signals with $T_1$ and $T_2$ values near the extremes of the parameter range (left and right sides of the graphs).

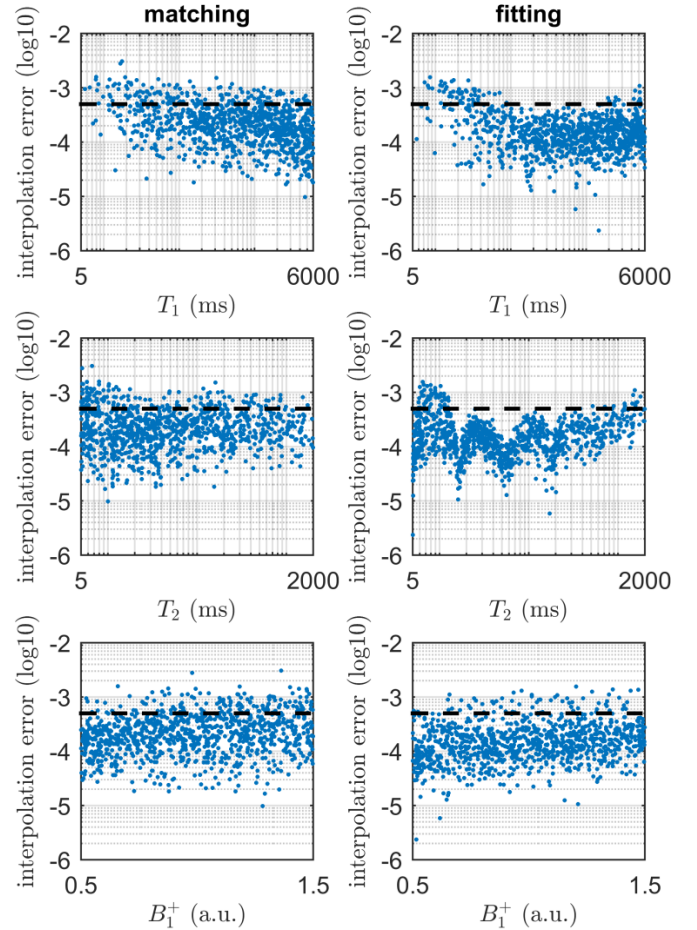

Fig. 2. Interpolation error at 1000 uniformly sampled positions in the grid for the dictionary used with matching (left) and fitting (right). The dashed line indicates the target error applied for dictionary design. The parameter resolution of both dictionaries is sufficient to obtain the target error for most grid positions in the interior.

A single evaluation of the spline interpolation function and its gradient took 1.4 ms without SVD compression.

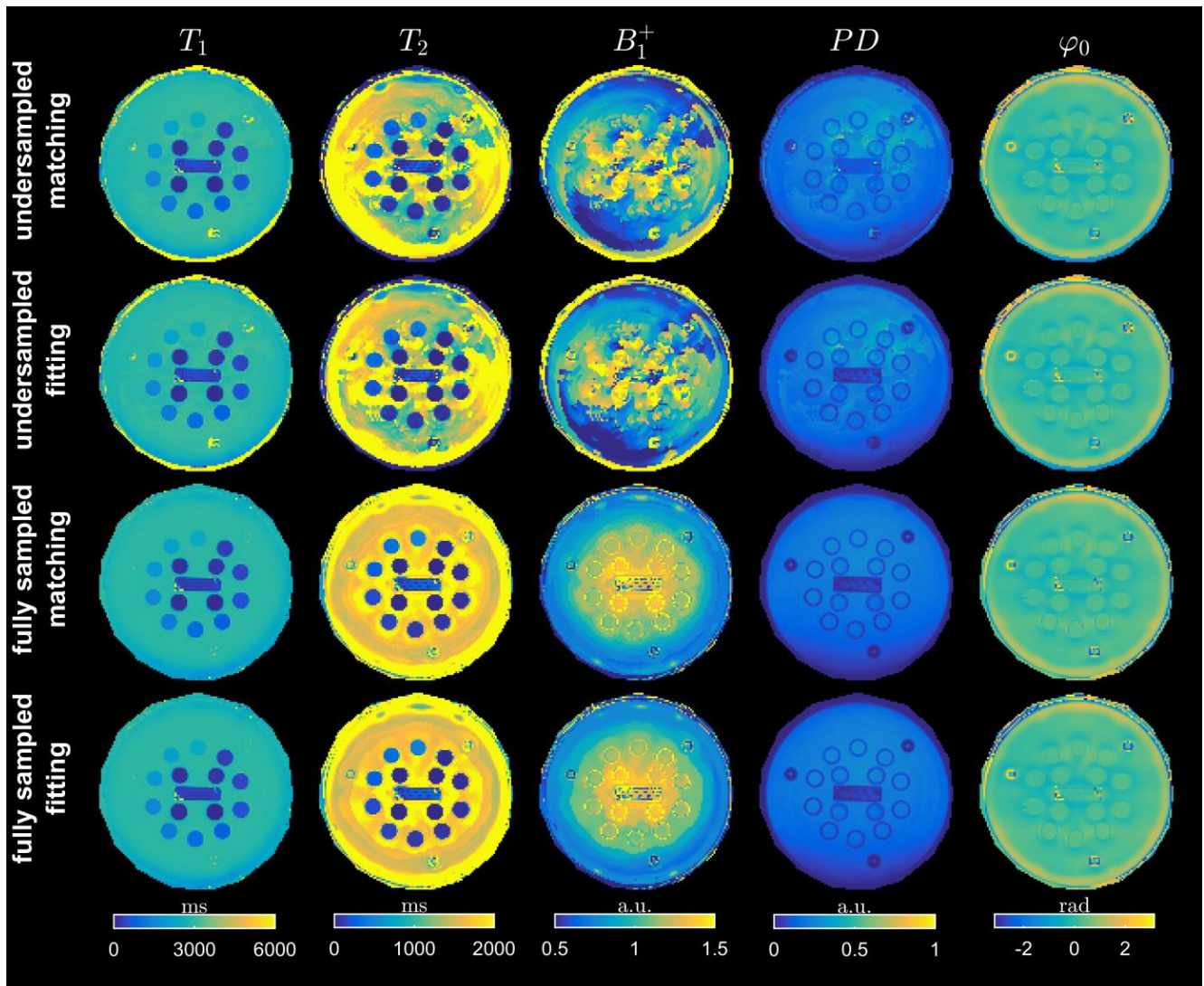

Fig. 3. Estimated parameter maps in the phantom through dictionary matching and dictionary fitting with both undersampled and fully sampled data. With both the undersampled data (top rows) as the fully sampled data (bottom rows), the dictionary matching maps are closely approximated by the proposed dictionary fitting method while only using 0.03% of the atoms.

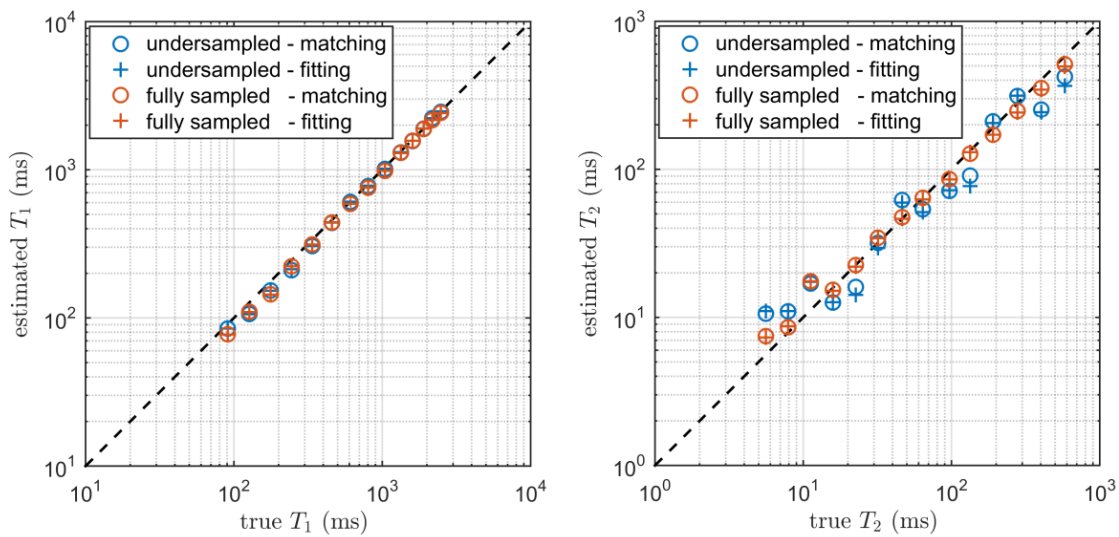

Fig. 4. Mean estimated value of $T_1$ (top) and $T_2$ (bottom) in each region-of-interest of the phantom as function of their calibrated value (log-log scale). Dictionary fitting obtains for both data sets and each parameter equal accuracy as dictionary matching while using 0.03% of the atoms.

## C. Phantom Experiment

Fig. 3 shows that both for the prospective undersampled and fully sampled acquisitions the parameter maps obtained with dictionary matching were closely approximated by the proposed dictionary fitting method. The $T_2$, $B_1^+$, and $PD$ maps from the undersampled data have some artefacts that are predominantly located in the background water.

Fig. 4 shows the mean estimated $T_1$ and $T_2$ values in each sphere of the phantom as a function of their calibrated values for undersampled and fully sampled acquisitions and both estimation methods. The relative differences between mean estimated and calibrated $T_1$ and $T_2$ values were respectively below 1.0% and 10.2% for the undersampled data, and below 0.7% and 3.1% for the fully sampled data. The root-mean-square error in the $T_1$ and $T_2$ estimates was similar for dictionary fitting and dictionary matching (see Supplementary Materials B).

The fitting time was 58 minutes for matching with the densely sampled dictionary while our proposed fitting method with the sparse dictionary took 6 minutes. These times did not include the loading of the dictionaries.

## D. In-vivo Experiment

Fig. 5 shows the parameter maps of the in-vivo experiment obtained from fully sampled data using both dictionary matching and dictionary fitting, as well as the difference between their maps. The parameter ranges of $T_1$ and $T_2$ are adjusted to highlight the tissues of interest. Differences between the two maps are mostly noticeable around the CSF, and both methods contain some residual structure in the $B_1$ map.

Compared to maps obtained without SVD compression, dictionary matching had a mean absolute relative error of 0.06% in $T_1$ and 1.32% in $T_2$, while dictionary fitting had an error of 0.15% in $T_1$ and 2.66% in $T_2$. The error of dictionary fitting was somewhat higher than dictionary matching, we hypothesize that the continuous optimization translates variation in the signal (due to the compression) directly to variation in the estimated parameters, while the discrete optimization requires significant variation in the signal before matching to another element of the dictionary and consequently another discretized parameter value.

The distribution of the error in the $T_1$ and $T_2$ maps as function of the number of sampled spirals is shown for both estimation methods in Fig. 6 (blue and red bars), where the error in each voxel is relative to the value obtained from the fully sampled data with the same estimation method. The proposed method has a smaller spread in error (indicated by the whiskers) than dictionary matching in most maps of each parameter except for the $T_2$ maps obtained from 1 and 6 spirals. Note that from 6 spirals onwards the dictionary matching approach selected the same atom as the fully sampled reference in the majority of voxels (boxes have zero width) and in most others one step in the dictionary away (whiskers), while the continuous estimate of the proposed fitting approach has a small but finite width. Furthermore, it can be noticed that the relative $T_1$ error was below the relative $T_2$ error.

## V. Discussion

This work presented a novel method for quantitative parameter estimation based on the least-squares fitting of a signal model defined by B-spline interpolation of a sparsely sampled dictionary. The FISP MRF sequence was chosen as the basis imaging sequence due to its ability to estimate multiple parameters simultaneously, though the precision of $T_1$ appears to be superior to that of $T_2$ for this sequence [8]. However, the proposed dictionary fitting framework is applicable for general acquisitions and parameters.

The interpolation error was estimated as a function of the parameter resolution for different B-splines orders. With second or third order splines, the resolution of each parameter reduced by approximately an order of magnitude compared to nearest neighbor interpolation. Consequently, the total number of atoms in the dictionary could be reduced with three orders of magnitude, leading to an equal reduction in memory and computational costs while maintaining equal signal accuracy.

The large reduction of resolution of each parameter makes it computationally feasible to estimate an increased number of parameters simultaneously. In Supplementary Materials C, we demonstrated this by constructing and fitting with a five-dimensional dictionary; additionally including intra-voxel dephasing $T_2'$ and off-resonance frequency $\Delta\omega_0$. This only increased the dictionary size by a factor 164 and fitting time by 58.4%. Consequently, the reduced computational and memory costs of dictionary-based methods enables the development of acquisition schemes that estimate more parameters simultaneously. Furthermore, the accuracy of our model can be increased by extending the Bloch simulation while using similar computing resources for the dictionary generation. Finally, the smaller dictionary sizes benefit methods that require dictionary generation on-the-fly, e.g. to incorporate acquisition details such as movement in the signal model [20].

The interpolation error of the simulated signals was found to be slightly higher than the predetermined threshold at some points near the boundary of the parameter domains. We performed an additional experiment (not shown) with interior points that only require one-dimensional interpolation (i.e. restricting the other two dimensions to the grid), and found that the number of interpolation errors above the threshold reduced to less than 1%, with a maximum of $6.8 \cdot 10^{-4}$. This shows that the error is predominantly caused by interpolating in multiple dimensions, while our parameter resolution was based on one-dimensional interpolation. A practical solution would be to set the actually applied threshold somewhat below the preferred accuracy (a factor two is appropriate for our three-dimensional dictionary). Furthermore, we assumed that the interpolation error is maximal near the boundary of the parameter range and Fig. 2 showed that this was the case for $T_1$ and $T_2$. However, this may not be true for each pulse sequence and for every parameter. Therefore, an evaluation of the interpolation error in the interior of the dictionary is recommended for general application.

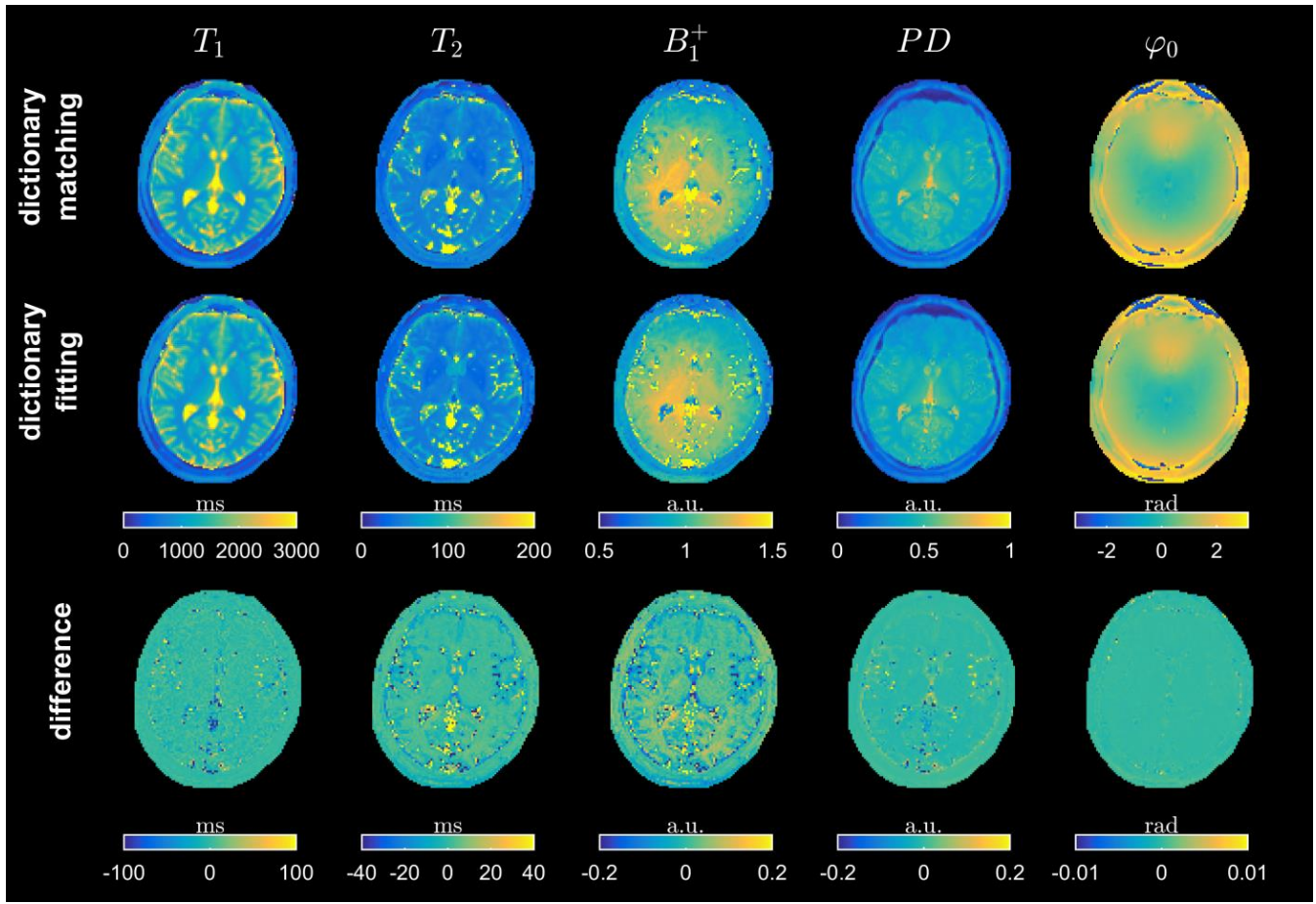

Fig. 5. Parameter maps in the brain from the in-vivo experiment, estimated through dictionary matching (top) and dictionary fitting (middle), and their difference (bottom). Note that the parameter ranges of the difference maps have been adjusted to highlight the differences.

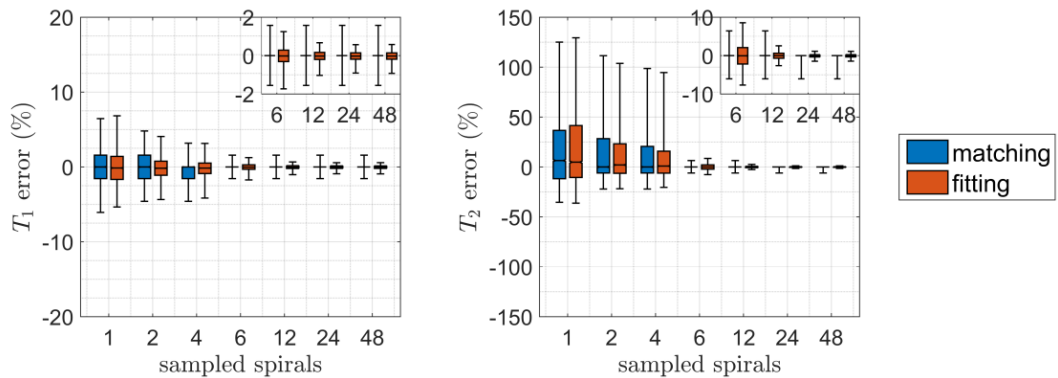

Fig. 6. Box-and-whisker plots of relative difference in $T_1$ (left) and $T_2$ (right) parameter values in the brain as function of the number of spirals/interleaves generated by prospective undersampling k-space. Boxes represent 25-75 percentiles and whiskers indicate the 5-95 percentiles of the error values of all voxels in the brain. Both estimation methods have a similar error spread in $T_1$ and $T_2$ for each number of sampled spirals.

The phantom and volunteer experiments showed that the proposed dictionary fitting method was able to estimate $T_1$ and $T_2$ with similar accuracy as dictionary matching while reducing the number of atoms three orders of magnitude. The in-vivo $B_1^+$ maps had residual structure near the CSF, which is probably due to correlation between the $B_1^+$ and $T_2$ parameters, which is known for MRF methods [12], [13].

In the presented results, we chose as error threshold $\alpha = 5 \cdot 10^{-4}$ and assumed that this was sufficiently accurate for errors in the parameters to be dominated by noise (and not e.g. by discretization errors). In Supplementary Materials D, we examined the quality of the in-vivo $T_1$ and $T_2$ maps when setting α a factor 10 higher and lower. This showed that the $T_1$ maps were reasonably consistent for different $\alpha$ and B-spline orders $n \in \{0,1,2,3\}$. The $T_2$ maps showed large variation with $\alpha = 5 \cdot 10^{-3}$, and small differences around the CSF for $\alpha = 5 \cdot 10^{-5}$. Thus, small improvements in $T_2$ estimation might be possible by lowering the threshold $\alpha$. However, we chose not to do this since the computational and memory requirements of the reference method would be too high for our available resources.

The proposed dictionary fitting method reduced the calculation time of the fit compared to the matching with a dense dictionary. The calculation time of both estimation techniques can be further reduced by parallelizing the fitting over multiple cores. Additionally, the dictionary matching can benefit from smarter search strategies [14], although application to higher dimensions is still limited due to the required dictionary size. The proposed dictionary fitting method used the `trust-region-reflective` algorithm since it was recommended by the MATLAB documentation for constrained optimization with gradients. While we experienced that convergence was reasonably fast, often within 20 iterations, further improvements can likely be found when doing an in-depth analysis of the applied solver. An alternative parameter estimation method is directly fitting the data to the Bloch equations. However, this would require a strong simplification of our signal model as generating only a single atom currently already took 6.10 seconds.

The proposed dictionary fitting method had similar accuracy as a dictionary matching strategy applying a dense dictionary, even when using data with undersampling artifacts. Hence, dictionary fitting is a beneficial substitute in many cases where dictionary matching is currently used. It can be directly inserted in iterative reconstruction methods with undersampled MRF data [20], by replacing the pattern matching with dictionary fitting. Initialization of the fitting through dictionary matching with the sparse dictionary likely enhances the probability of starting the optimization close to the global optimum.

The dictionary fitting framework was presented for a single MRF pulse sequences and associated model parameters, but is easily extendable to other qMRI methods as presented in [3]–[7].

## VI. CONCLUSION

The Bloch simulated signal is accurately and efficiently approximated through B-spline interpolation of a sparsely sampled dictionary. Therefore, the proposed method enables estimating parameters by fitting a continuous B-spline signal model, which obtains the accuracy of dictionary matching while strongly reducing dictionary size. The required parameter resolution is efficiently determined on the boundary of the parameter range. The proposed methods were applied to a FISP MRF acquisition in this work, but can be used for any qMRI acquisition scheme.

ACKNOWLEDGMENT

The authors thank Holland PTC for the use of their NIST system phantom.

REFERENCES

[1] H. L. M. Cheng, N. Stikov, N. R. Ghugre, and G. A. Wright, "Practical medical applications of quantitative MR relaxometry," *J. Magn. Reson. Imaging*, vol. 36, no. 4, pp. 805–824, Oct. 2012.

[2] D. Ma, V. Gulani, N. Seiberlich, K. Liu, J. L. Sunshine, J. L. Duerk, and M. A. Griswold, "Magnetic resonance fingerprinting," *Nature*, vol. 495, no. 7440, pp. 187–192, Mar. 2013.

[3] J. K. Barral, E. Gudmundson, N. Stikov, M. Etezadi-Amoli, P. Stoica, and D. G. Nishimura, "A robust methodology for in vivo T1 mapping," *Magn. Reson. Med.*, vol. 64, no. 4, pp. 1057–1067, Oct. 2010.

[4] J. D. Trzasko, P. M. Mostardi, S. J. Riederer, and A. Manduca, "Estimating T1 from multichannel variable flip angle SPGR sequences," *Magn. Reson. Med.*, vol. 69, no. 6, pp. 1787–1794, Jun. 2013.

[5] N. Ben-Eliezer, D. K. Sodickson, and K. T. Block, "Rapid and accurate T2 mapping from multi–spin-echo data using Bloch-simulation-based reconstruction," *Magn. Reson. Med.*, vol. 73, no. 2, pp. 809–817, Feb. 2015.

[6] D. Hernando, J. P. Haldar, B. P. Sutton, J. Ma, P. Kellman, and Z.-P. Liang, "Joint estimation of water/fat images and field inhomogeneity map," *Magn. Reson. Med.*, vol. 59, no. 3, pp. 571–580, Mar. 2008.

[7] D. Hernando, P. Kellman, J. P. Haldar, and Z.-P. Liang, "Robust water/fat separation in the presence of large field inhomogeneities using a graph cut algorithm," *Magn. Reson. Med.*, vol. 63, no. 1, pp. 79–90, Jan. 2010.

[8] Y. Jiang, D. Ma, N. Seiberlich, V. Gulani, and M. A. Griswold, "MR fingerprinting using fast imaging with steady state precession (FISP) with spiral readout," *Magn. Reson. Med.*, vol. 74, no. 6, pp. 1621–1631, Dec. 2015.

[9] Y. Jiang, D. Ma, K. Wright, N. Seiberlich, V. Gulani, and M. A. Griswold, "Simultaneous T1, T2, diffusion and proton density quantification with MR fingerprinting," in *Proc. ISMRM*, Milan, Italy, 2014.

[10] P. Su, D. Mao, P. Liu, Y. Li, M. C. Pinho, B. G. Welch, and H. Lu, "Multiparametric estimation of brain hemodynamics with MR fingerprinting ASL," *Magn. Reson. Med.*, vol. 78, no. 5, pp. 1812–1823, 2017.

[11] J. I. Hamilton and N. Seiberlich, "MR fingerprinting with chemical exchange (MRF-X) for in vivo multi-compartment relaxation and exchange rate mapping," in *Proc. ISMRM*, Singapore, 2016.

[12] G. Buonincontri and S. J. Sawiak, "MR fingerprinting with simultaneous B1 estimation," *Magn. Reson. Med.*, vol. 76, no. 4, pp. 1127–1135, Oct. 2016.

[13] D. Ma, S. Coppo, Y. Chen, D. F. McGivney, Y. Jiang, S. Pahwa, V. Gulani, and M. A. Griswold, "Slice profile and B1 corrections in 2D magnetic resonance fingerprinting," *Magn. Reson. Med.*, vol. 78, no. 5, pp. 1781–1789, 2017.

[14] S. F. Cauley, K. Setsompop, D. Ma, Y. Jiang, H. Ye, E. Adalsteinsson, M. A. Griswold, and L. L. Wald, "Fast group matching for MR fingerprinting reconstruction," *Magn. Reson. Med.*, vol. 74, no. 2, pp. 523–528, 2015.

[15] D. F. McGivney, E. Pierre, D. Ma, Y. Jiang, H. Saybasili, V. Gulani, and M. A. Griswold, "SVD Compression for Magnetic Resonance Fingerprinting in the Time Domain," *IEEE Trans. Med. Imaging*, vol. 33, no. 12, pp. 2311–2322, Dec. 2014.

[16] M. Yang, D. Ma, Y. Jiang, J. Hamilton, N. Seiberlich, M. A. Griswold, and D. McGivney, "Low rank approximation methods for MR fingerprinting with large scale dictionaries," *Magn. Reson. Med.*, Aug. 2017.

[17] M. Unser, "Splines: A perfect fit for signal and image processing," *IEEE Signal Process. Mag.*, vol. 16, no. 6, pp. 22–38, 1999.

[18] W. Huizinga, S. Klein, and D. H. J. Poot, "Fast Multidimensional B-spline Interpolation Using Template Metaprogramming," in *WBIR*, 2014, pp. 11–20.

[19] K. E. Keenan, K. F. Stupic, M. A. Boss, S. E. Russek, T. L. Chenevert, P. V. Prasad, W. E. Reddick, J. Zheng, P. Hu, and E. F. Jackson, "Comparison of T1 measurement using ISMRM/NIST system phantom," in *Proc. ISMRM*, Singapore, 2016.

[20] J. I. Hamilton, Y. Jiang, Y. Chen, D. Ma, W.-C. Lo, M. Griswold, and N. Seiberlich, "MR fingerprinting for rapid quantification of myocardial T1, T2, and proton spin density," *Magn. Reson. Med.*, vol. 77, no. 4, pp. 1446–1458, Apr. 2017.

# An Efficient Method for Multi-Parameter Mapping in Quantitative MRI using B-Spline Interpolation

Willem van Valenberg, Stefan Klein, Frans M. Vos, Kirsten Koolstra, Lucas J. van Vliet, Dirk H.J. Poot

## A. EQUIVALENT OPTIMIZATION

Here we show that solution $\hat{\boldsymbol{\theta}} = \boldsymbol{f}(\hat{\boldsymbol{k}})$ of the dictionary matching step in Eq. 3 with the condition $\hat{\rho} = \left(\boldsymbol{s}(\hat{\boldsymbol{\theta}})^H \boldsymbol{m}\right) / \left(\boldsymbol{s}(\hat{\boldsymbol{\theta}})^H \boldsymbol{s}(\hat{\boldsymbol{\theta}})\right)$ as given in Eq. 4, is also the solution of the optimization in Eq. 2 when restricting the search space $\Theta$ to the parameter combinations in the dictionary. Setting $\rho = \rho_R + i\rho_I$, we can write the error term in Eq. 2 as function of $\rho_R, \rho_I \in \mathbb{R}$:

$$\|\boldsymbol{m} - \rho \boldsymbol{s}(\boldsymbol{\theta})\|_2^2 = \left(\boldsymbol{m} - \rho \boldsymbol{s}(\boldsymbol{\theta})\right)^H \left(\boldsymbol{m} - \rho \boldsymbol{s}(\boldsymbol{\theta})\right) = \left(\overline{\boldsymbol{m}} - \rho_R \overline{\boldsymbol{s}(\boldsymbol{\theta})} + i\rho_I \overline{\boldsymbol{s}(\boldsymbol{\theta})}\right)^T \left(\boldsymbol{m} - \rho_R \boldsymbol{s}(\boldsymbol{\theta}) - i\rho_I \boldsymbol{s}(\boldsymbol{\theta})\right)$$

The bars indicate complex conjugation. Setting the partial derivative with respect to $\rho_R$ or $\rho_I$ to zero, gives for both cases Eq. 4 as necessary condition for the minimum. With condition Eq. 4, the error term in Eq. 2 can be written as:

$$\|\boldsymbol{m} - \hat{\rho}\boldsymbol{s}(\boldsymbol{\theta})\|_2^2 = \|\boldsymbol{m}\|_2^2 + |\hat{\rho}|^2 \|\boldsymbol{s}(\boldsymbol{\theta})\|_2^2 - \boldsymbol{m}^H \hat{\rho} \boldsymbol{s}(\boldsymbol{\theta}) - \hat{\rho}^H \boldsymbol{s}(\boldsymbol{\theta})^H \boldsymbol{m} = \|\boldsymbol{m}\|_2^2 + \frac{|\boldsymbol{s}(\boldsymbol{\theta})^H \boldsymbol{m}|^2}{\|\boldsymbol{s}(\boldsymbol{\theta})\|_2^2} - 2\frac{\boldsymbol{m}^H \boldsymbol{s}(\boldsymbol{\theta}) \cdot \boldsymbol{s}(\boldsymbol{\theta})^H \boldsymbol{m}}{\|\boldsymbol{s}(\boldsymbol{\theta})\|_2^2}$$

$$= \|\boldsymbol{m}\|_2^2 - \frac{|\boldsymbol{m}^H \boldsymbol{s}(\boldsymbol{\theta})|^2}{\|\boldsymbol{s}(\boldsymbol{\theta})\|_2^2}$$

Since $\boldsymbol{m}$ is fixed, the minimization in Eq. 2 is equal to the maximization in Eq. 3 over the discrete parameter values $\boldsymbol{f}(\boldsymbol{k})$.

## B. ERROR IN $T_1$ AND $T_2$ ESTIMATES FROM PHANTOM EXPERIMENT

Table S1 shows the root-mean-square error (RMSE) in the estimated values of $T_1$ and $T_2$ from the phantom experiment, relative to their calibrated values. For most cases, the proposed dictionary fitting method reduced the error in the parameter values compared to the reference dictionary matching method. Exceptions were mostly found in the $T_2$ estimation, where dictionary matching was more accurate than fitting in ROIs 1-3, and more precise in ROIs 7 and 8. The reduced accuracy of the proposed method for high $T_2$ values (ROIs 1-3) was possibly due to logarithmic spacing of the parameter in the dictionary, leading to large steps between high $T_2$ values and consequently an inaccurate initialization of the fit. The higher precision of dictionary matching in ROIs 7 and 8 might be due to the matching of the voxels in each ROI to a discrete set of $T_2$ values, which can lower the variance if (almost) all voxels are matched to the same parameter value. Overall, the approximation of the signal model through spline interpolation did not increase the error in the estimated $T_1$ and $T_2$ compared to dictionary matching.

Table S1 Root-mean-square error (RMSE) of the estimated $T_1$ and $T_2$ values in the voxels of each region-of-interest (ROI) of the phantom with respect to the calibrated value. The error is given as percentage of the calibrated value in each ROI.

| ROI | calibrated $T_1$ [ms] | RMSE in $T_1$ [%] | | | | calibrated $T_2$ [ms] | RMSE in $T_2$ [%] | | | |
|---|---|---|---|---|---|---|---|---|---|---|
| | | undersampled | | fully sampled | | | undersampled | | fully sampled | |
| | | matching | fitting | matching | fitting | | matching | fitting | matching | fitting |
| 1 | 2480 | 1.8 | 1.1 | 2.6 | 2.7 | 581 | 31.4 | 40.8 | 12.6 | 15.4 |
| 2 | 2173 | 3.0 | 2.7 | 0.9 | 0.8 | 404 | 42.7 | 42.7 | 14.1 | 15.9 |
| 3 | 1907 | 2.6 | 2.6 | 2.0 | 1.7 | 278 | 17.4 | 22.3 | 13.9 | 15.6 |
| 4 | 1604 | 3.4 | 3.2 | 2.8 | 2.7 | 191 | 34.0 | 31.2 | 13.2 | 12.7 |
| 5 | 1332 | 3.6 | 3.1 | 3.3 | 3.5 | 133 | 52.2 | 49.9 | 9.4 | 7.3 |
| 6 | 1044 | 4.5 | 4.3 | 6.8 | 6.5 | 97 | 32.1 | 30.8 | 13.7 | 14.5 |
| 7 | 802 | 5.1 | 5.0 | 6.6 | 6.4 | 64 | 27.3 | 31.7 | 7.9 | 12.0 |
| 8 | 609 | 5.5 | 5.3 | 5.2 | 5.2 | 46 | 60.2 | 59.4 | 9.0 | 19.1 |
| 9 | 458 | 7.0 | 6.7 | 4.9 | 4.7 | 32 | 40.6 | 37.5 | 12.5 | 10.8 |
| 10 | 337 | 10.6 | 10.4 | 7.9 | 8.1 | 23 | 55.3 | 51.9 | 22.6 | 20.7 |
| 11 | 244 | 16.6 | 16.1 | 9.5 | 9.3 | 16 | 49.6 | 50.1 | 29.0 | 28.2 |
| 12 | 177 | 53.6 | 52.6 | 48.3 | 48.2 | 11 | 395.3 | 418.0 | 393.2 | 396.7 |
| 13 | 127 | 22.1 | 22.1 | 14.2 | 15.1 | 8 | 78.3 | 79.8 | 49.6 | 51.5 |
| 14 | 91 | 27.6 | 26.8 | 16.0 | 16.3 | 6 | 147.7 | 156.0 | 65.8 | 64.2 |

## C. HIGHER DIMENSIONAL PARAMETER ESTIMATION

Here we show that dictionary fitting enables the estimation of an increased number of parameters compared to dictionary matching. We added estimation of intra-voxel dephasing of spins $T_2'$ and static field inhomogeneity $\Delta B_0$ from the in-vivo scan, and consequently increased dimensionality of our dictionary from three to five. The number of spins used in our model is increased by a factor 10 in order to accurately model the intra-voxel dephasing.

Figure S1 shows the number of discretized values in each parameter dimension for B-spline orders $n = 0,1,2,3$, as determined through the method described in Section II.D. The interpolation error in the $T_2'$ and $\Delta \omega_0$ dimension converged to a value just below the target error. We found (data not shown) that the minimal interpolation error in these dimensions scaled with the square root of the number of spins used in the simulations. Dictionary fitting (Eq. 8) was done with B-spline order $n = 2$, using a dictionary with 11, 6, 5, 24, and 16 parameter values in respectively the ranges of $T_1, T_2, T_2', \Delta\omega_0$, and $B_1^+$. Storage of this dictionary required 2,34 B without SVD and 74,1 MB when projected on the first 30 singular vectors, a factor 164 increase compared to the three-dimensional dictionary. Fitting the in-vivo maps with the SVD-projected dictionaries took 301 seconds which is only 58,4% longer than the fitting time using the three-dimensional dictionary.

Figure S2 shows the parameter maps obtained from fitting with a dictionary with five parameter dimensions. The $T_2'$ parameter map obtained with the five-dimensional dictionary looks implausible, which is expected since the used acquisition is typically not used to estimate this parameter. The $\Delta\omega_0$ map shows some off-resonance around the sinuses which is likely due to the air-tissue interface, even though the applied acquisition is designed to be robust against off-resonance effects. The low signal region near the skull leads to the mapping of very low $T_2'$ values which also led to implausible values of $T_2, \Delta\omega_0$, and $B_1^+$.

Although for the evaluated FISP MRF acquisition the additional parameter maps were either implausible ($T_2'$) or typically irrelevant ($\Delta\omega_0$), we have shown that the dictionary fitting method is able to estimate seven parameters simultaneously, using a five-dimensional dictionary. If we suppose that for B-spline order $n = 0$, the target interpolation error in Figure S1 is obtained using 129 atoms for $T_1, T_2', \omega_0$ and $B_1^+$, then matching with a dictionary that obtains the target interpolation error would require 189 TB storage for the dictionary (without SVD), and $4.8 \cdot 10^7$ $(= 100 \cdot 129^4 \cdot \frac{6.2}{3600})$ computer hours to generate (with the used Bloch simulation that requires on average 6.2 s per atom). The proposed method enables the design of new quantitative acquisitions that can estimate an increased number of parameters simultaneously and consequently obtain more biomarkers (e.g. $T_1, T_2, T_2', PD$) with reduced bias due to the magnetic fields ($\Delta\omega_0, B_1^+, \varphi_0$).

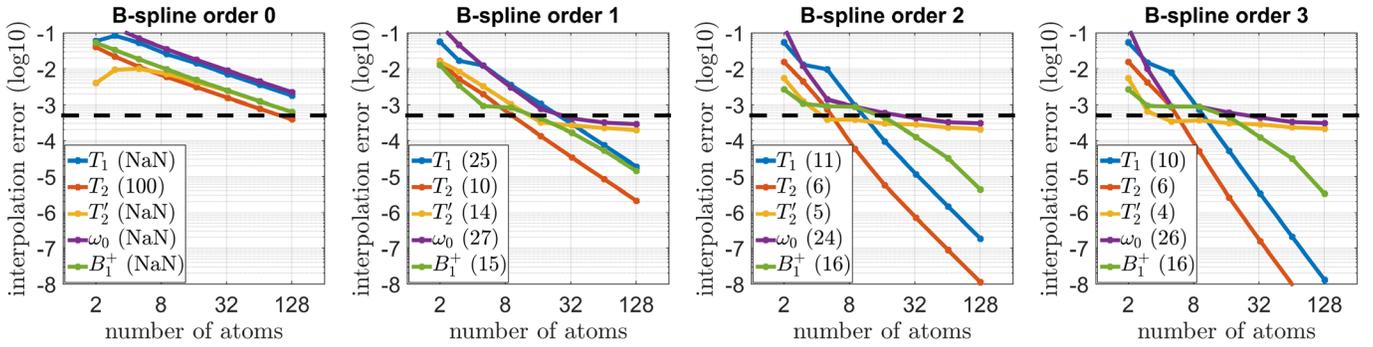

Fig. S1 Interpolation error on the edge of parameter space as function of number of atoms in each dimension. Dashed line indicates target error and the legend shows for each parameter the minimum number of atoms required to obtain the target error. Note that for B-spline order $n = 0$, the target interpolation error is only obtained in the $T_2$ dimension with less than 130 atoms.

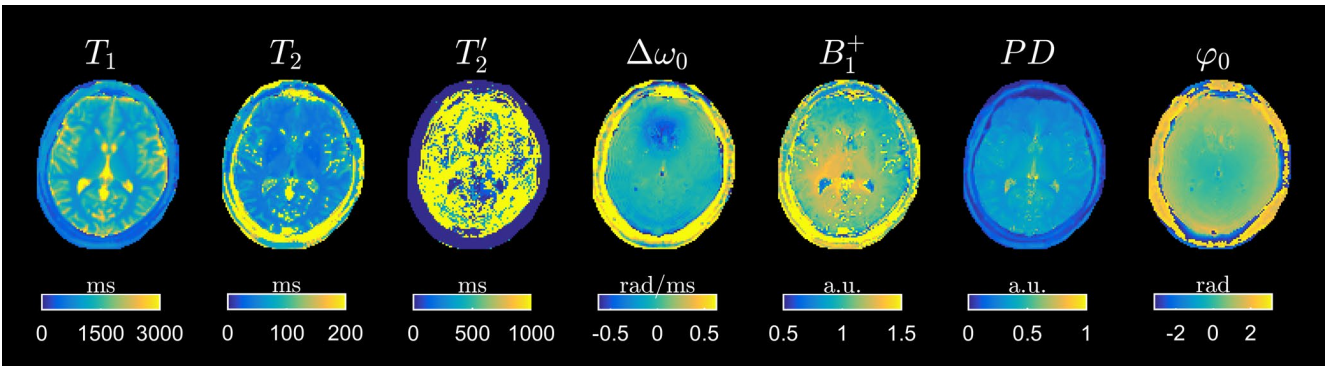

Fig. S2 Parameter maps in the brain from dictionary fitting with B-spline order $n = 2$ using a five-dimensional dictionary. Although the $T_2'$ and $\Delta\omega_0$ maps have low quality, this example shows that the proposed method is computationally feasible for seven-dimensional parameter estimation.

## D. EFFECT OF B-SPLINE ORDER AND INTERPOLATION ACCURACY

Here we present a qualitative comparison of the $T_1$ and $T_2$ parameter maps obtained through dictionary fitting with different combinations of B-spline order $n \in \{0,1,2,3\}$ and interpolation accuracy $\alpha \in \{5 \cdot 10^{-3}, 5 \cdot 10^{-4}, 5 \cdot 10^{-5}\}$. For each combination, we generated a dictionary with parameter resolution as predicted by the methods described in Section II.D (i.e. by shifting the threshold in each graph of Fig. 1). The resulting parameter resolutions are shown in Table S2. The combination $n = 0$ with $\alpha = 5 \cdot 10^{-5}$ was excluded since it required more than the maximal number (512) of atoms in each parameter dimension. This would require more than a TB memory for dictionary storage, which is unfeasible for our available resources.

Fig. S3a shows the $T_1$ maps from dictionary fitting with each combination of B-spline order $n$ and interpolation threshold $\alpha$. The different $T_1$ maps appear reasonably consistent, except for $n = 0$ with $\alpha = 5 \cdot 10^{-3}$ where some quantization error is observable. Fig. S3b contains the $T_2$ maps from the same combinations of $n$ and $\alpha$. Here, the differences between the maps are larger. There is a large variation between the $T_2$ maps with interpolation error $\alpha = 5 \cdot 10^{-3}$, which suggests that this error threshold is too high for reliably estimating this parameter. The maps with threshold $\alpha = 5 \cdot 10^{-4}$ and $\alpha = 5 \cdot 10^{-5}$ are more consistent and only show some slight variation around the CSF.

Table S2 Estimated number of atoms required in each parameter dimension to obtain a given interpolation accuracy $\alpha$ using B-spline order $n$. These numbers were estimated using the method described in Section II.D.

|  | $n = 0$ | | | $n = 1$ | | | $n = 2$ | | | $n = 3$ | | |
|---|---|---|---|---|---|---|---|---|---|---|---|---|
|  | $T_1$ | $T_2$ | $B_1^+$ | $T_1$ | $T_2$ | $B_1^+$ | $T_1$ | $T_2$ | $B_1^+$ | $T_1$ | $T_2$ | $B_1^+$ |
| $\alpha = 5 \cdot 10^{-3}$ | 47 | 11 | 17 | 8 | 3 | 2 | 6 | 3 | 2 | 6 | 3 | 2 |
| $\alpha = 5 \cdot 10^{-4}$ | 454 | 97 | 160 | 25 | 10 | 17 | 11 | 6 | 18 | 10 | 6 | 18 |
| $\alpha = 5 \cdot 10^{-5}$ | | $> 512$ | | 80 | 27 | 98 | 21 | 10 | 78 | 18 | 9 | 77 |

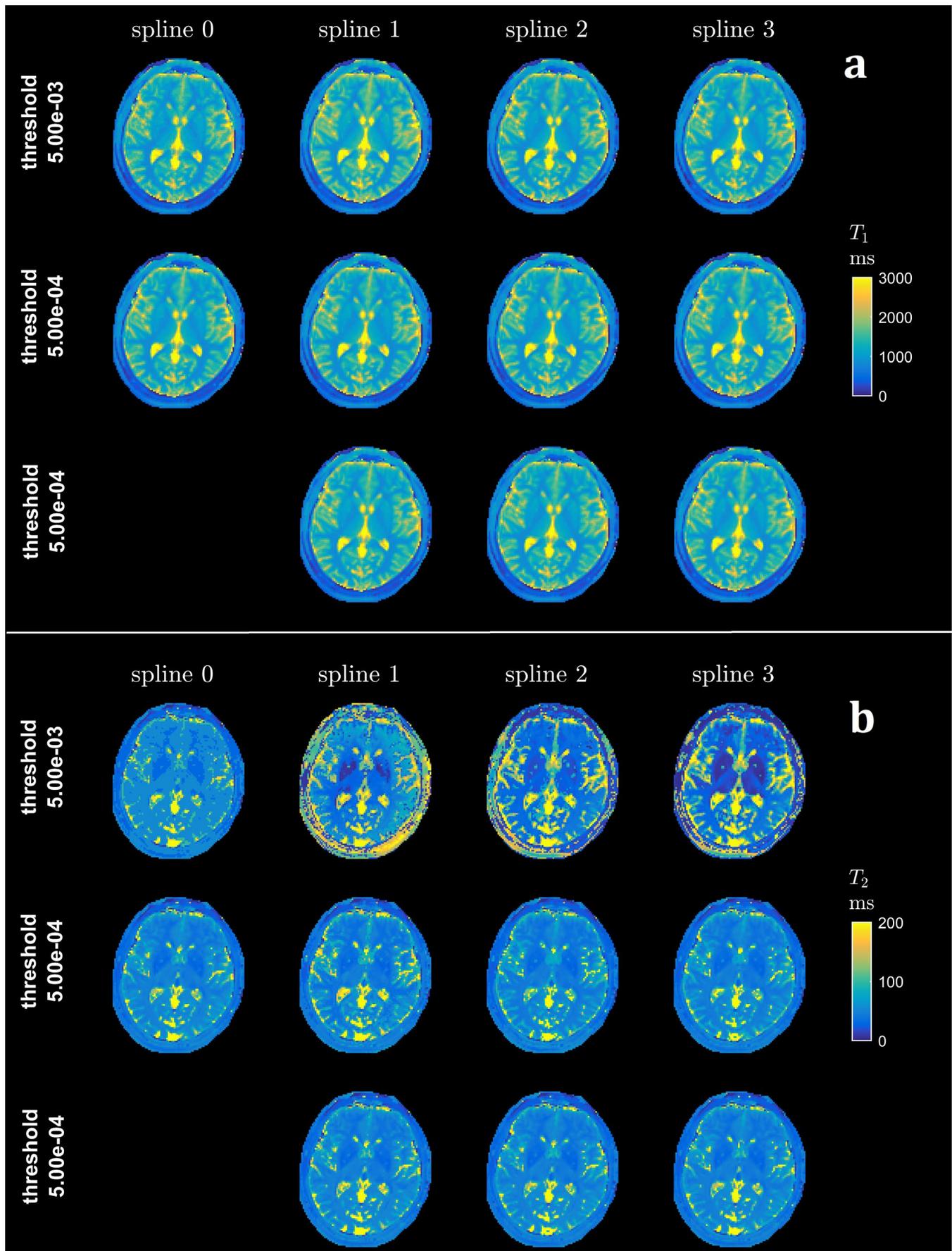

Fig. S3 Parameter maps of $T_1$ (a) and $T_2$ (b) in the brain for B-spline orders $n = 0,1,2,3$ (columns) and for each B-spline order multiple levels of interpolation accuracy $\alpha$. The combination $n = 0$ and $\alpha = 5 \cdot 10^{-4}$ was not included since the required dictionary size was too large for our available resources.